\documentclass[prb,amsmath,superscriptaddress,citeautoscript,twocolumn,showpacs,floatfix]{revtex4}
\usepackage{graphicx}

\usepackage{amssymb}
\usepackage{color}
\usepackage{epsfig}

\def\sig{{\mbox{\boldmath{$\sigma$}}}}

\def\b0{{\bf{0}}}

\begin{document}

\def\sigmav{{\mbox{\boldmath{$\sigma$}}}}

\title{Partial decoherence in mesoscopic systems}
\author{Amnon Aharony}
\email{aaharony@bgu.ac.il}
\altaffiliation{Also at Tel Aviv University, Tel Aviv 69978,
Israel}

\affiliation{Department of Physics and the Ilse Katz Center for
Meso- and Nano-Scale Science and Technology, Ben-Gurion University, Beer
Sheva 84105, Israel}

\author{Shmuel Gurvitz}
\affiliation{Department of Particle Physics and Astrophysics,
Weizmann Institute of Science, Rehovot 76100, Israel}

\author{Yasuhiro Tokura}
\altaffiliation{Present address: Institute of Physics, University of Tsukuba, Tsukuba, 305-1571, Japan}

\affiliation{ NTT Basic Research Laboratories, NTT Corporation,
Atsugi-shi, Kanagawa 243-0198, Japan}
\author{Ora Entin-Wohlman}
\altaffiliation{Also at Tel Aviv University, Tel Aviv 69978,
Israel}

\affiliation{Department of Physics and the Ilse Katz Center for
Meso- and Nano-Scale Science and Technology, Ben-Gurion
University, Beer Sheva 84105, Israel}

\author{Sushanta Dattagupta}
\affiliation{ Indian Institute of Science Education and
Research-Kolkata, Mohanpur 741252, India}

\date{\today}
\begin{abstract}

The coupling of a mesoscopic system with its environment usually causes total decoherence: at long times the reduced density matrix of the system evolves in time to a limit which is independent of its initial value, losing all the quantum information stored in its initial state.  Under special circumstances, a subspace of the system's Hilbert space remains coherent, or ``decoherence free", and the reduced density matrix approaches a non-trivial limit which contains  information on its initial quantum state, despite the coupling to the environment. This situation is called ``partial decoherence". Here we find the conditions for partial decoherence for a mesoscopic system (with $N$ quantum states) which is coupled to an environment. When the Hamiltonian of the system commutes with the total Hamiltonian, one has ``adiabatic decoherence", which yields  $N-1$ time-independent combinations of the reduced density matrix elements. In the presence of a magnetic flux, one can measure circulating currents around loops in the system even at long times, and use them to retrieve information on the initial state. For $N=2$, we demonstrate that partial decoherence can happen only under adiabatic decoherence conditions. However, for $N>2$ we find partial decoherence even when the Hamiltonian of the system does not commute with the total Hamiltonian, and we obtain the general conditions for such non-adiabatic partial decoherence. For an electron moving on a ring, with $N>2$ single-level quantum dots,  non-adiabatic partial decoherence can arise only when the total flux through the ring vanishes (or equals an integer number of flux quanta), and therefore  there is no asymptotic circulating current.

\end{abstract}

\maketitle

\section{Introduction}

Quantum computation operates on information  stored in ``qubits",
which are  superpositions of two basic quantum states.
\cite{bennett}   Clearly, quantum computation requires the
stability of the quantum state stored on each qubit, and therefore
it can be used only while this state remains coherent
\cite{zurek}. Interactions between  qubits and their environment,
including input-output measurement devices, can cause decoherence
which destroys the information stored in the qubits. Attempts to
avoid decoherence have led to studies of decoherence-free
subspaces, within which the quantum state remains protected
\cite{Lidar,Zanardi}. The states in such subspaces are practically
decoupled from the sources of the decoherence, due to symmetries
of the system. The existence of such states means that the system has only {\it partial decoherence}.

Consider a qubit which is based on two single-level quantum
dots.\cite{divin} Denoting the states on each dot by
 $|1\rangle$ and $|2\rangle$, the initial state of the qubit is written as
\begin{align}\label{psi0}
|\psi^{}_0\rangle=\cos\alpha|1\rangle+e^{i\gamma}\sin\alpha
|2\rangle\ ,
\end{align}
with two real parameters $\alpha$ and $\gamma$. In a recent paper,\cite{prb} some of us considered
the coupling of such a two-quantum-dot system to a general environment, and found conditions for having partial decoherence.
An appropriate tuning of the dot energies and the coupling energies between the two dots yielded conditions under which one is able to retrieve the quantum information stored on the dot even after a long time, despite the decoherence.

An alternative system involves coupling the two-dot qubit to a third dot, $|3\rangle$, which acts as a ``control
register".\cite{divin} Only the latter dot is coupled to the environment. Such a ring of three dots was considered in  proposals \cite{kulik} to reduce the decoherence of
the single qubit state.
Indeed, an analysis of the time evolution of the three-site ring, with a coupling of the `control register' to an external fluctuator,  \cite{condmat}
has found partial decoherence under appropriate conditions.

The above two systems are special cases of a mesoscopic system, which contains $N$ single-level quantum dots (or any other system which has $N$ quantum states).
In this paper we generalize the discussion of Ref.$\ $\onlinecite{prb}, and study the conditions under which such a system can exhibit partial decoherence.
 To study decoherence, we consider the time evolution of the system
 in the
presence of the environment. To concentrate on the state of the mesoscopic
system, one traces the total density matrix of the system plus the environment
over the states of the latter, ending up
with the $N\times N$ {\it reduced density matrix} of the system itself,
\begin{align}
\rho(t)\equiv{\rm Tr}^{}_{\rm
env}[|\Psi(t)\rangle\langle\Psi(t)|]\ ,
\end{align}
 where $|\Psi\rangle$ is the
combined state of the system and the environment.

 In this paper we consider the following Hamiltonian:
\begin{align}
{\cal H}={\cal H}^{}_0+{\cal H}^{}_{\rm env}+{\cal V}{\cal U}^{}_{\rm env}\ ,\label{HHH}
\end{align}
where the first two terms describe the separate system and environment, and the last term describes the coupling between them. Here, the operator ${\cal V}$ acts on the system, while the operator ${\cal U}^{}_{\rm env}$ acts on the environment. Examples of such couplings include the coupling of the system to a vibrational mode, e.g. ${\cal V}{\cal U}^{}_{\rm env}=\zeta|1\rangle\langle 1|(b+b^\dagger)$, where $b^\dagger$ creates an excitation of this mode,\cite{phonon} or to electrons on a neighboring single-electron transistor, ${\cal V}{\cal U}^{}_{\rm env}=\zeta|1\rangle\langle 1|c^\dagger c$, where $c^\dagger$ creates an electron on the transistor.\cite{SET}

It is convenient to work with the eigenstates $|\ell\rangle$
of ${\cal H}^{}_0$,  with eigen-energies
$\tilde{\epsilon}^{}_\ell$:
\begin{align}
{\cal H}^{}_0|\ell\rangle=\tilde{\epsilon}^{}_\ell|\ell\rangle\ .\label{eig0}
\end{align}
From now on, we denote these eigenstates by $\ell$ or $\ell'$, and the states localized on specific quantum dots by $n$ or $m$.  If the system is decoupled from the
environment (i.e. ${\cal V}=0$), then the off-diagonal density
matrix elements (in the basis of the $|\ell\rangle$'s) exhibit
Rabi oscillations,
\begin{align}\label{rabi}
\rho^{}_{\ell\ell'}(t)=e^{i(\tilde{\epsilon}^{}_{\ell'}-\tilde{\epsilon}^{}_\ell)t}\rho^{}_{\ell\ell'}(0)
\end{align}
(we use $\hbar=1$).
 However,
 the coupling to the environment, reflected by the last term in Eq. (\ref{HHH}), implies that transitions between environment states can reduce the coherence of the system states.\cite{stern} Such decoherence usually causes the decay of all the off-diagonal reduced density matrix elements to zero. This decay is often attributed to random fluctuations of the phases of these matrix elements, and therefore it is called ``{\it dephasing}".  Although this decay has been explicitly demonstrated in many special cases [see e.g. Refs.$\ $\onlinecite{prb,condmat,SET,16,privman,17}], we are not aware of its general proof  for an arbitrary environment.

 Without special tuning, the diagonal reduced density matrix elements also approach an asymptotic limit which is independent of the initial quantum state. For example, if the environment is a heat bath at temperature $T$, the system usually {\it relaxes}, or {\it thermalizes},  towards thermal equilibrium, $\rho^{}_{\ell\ell'}(t\rightarrow\infty)=\delta^{}_{\ell\ell'}e^{-\tilde{\epsilon}^{}_\ell/(kT)}/\sum_\ell[e^{-\tilde{\epsilon}^{}_\ell/(kT)}]$.  At $T=0$ this leaves only the ground state, while at $T\rightarrow\infty$ this yields the {\it fully mixed} state,  $\rho^{}_{\ell\ell}(t\rightarrow\infty)=1/N$.  Again, although this behavior is found in many numerical or approximate calculations, we are not aware of any exact statements proving it.

 Below we discuss situations, in which some diagonal matrix elements do not evolve towards these trivial limits. Instead, they remain time-invariant, yielding {\it partial decoherence}. 
A special case of such partial decoherence has already been discussed in the literature.\cite{16,privman,17} If $[{\cal H}^{}_0,{\cal V}]=0$, then also $[{\cal H}^{}_0,{\cal H}]=0$, and the corresponding Heisenberg operator ${\cal H}^{}_0(t)$ becomes time independent. This situation has been called "{\it adiabatic decoherence}"\cite{16,privman}, or "{\it dissipationless decoherence}"\cite{17}. It has been shown\cite{16,privman} that in this case all the diagonal reduced density matrix elements $\rho^{}_{\ell\ell}$ remain time-invariant.

In the present paper we discuss the more general case, of {\it non-adiabatic partial decoherence}, when $[{\cal H}^{}_0,{\cal V}]\ne 0$. Even without this commutator, we find that some parts of the initial state information can be retrieved from the reduced density matrix at any time.
Adiabatic decoherence then turns out to be a special case of our general discussion, and we show that this is the {\it only} case which yields partial decoherence for $N=2$.\cite{prb}  However,  for $N>2$ one can also have  other situations.

The general conditions for partial decoherence are discussed in
Sec. II.   Section III demonstrates these conditions for a few examples.
Generally, we consider a tight-binding Hamiltonian, with ``site" energies $\epsilon^{}_n$ and ``hopping"
matrix elements $J^{}_{nm}$,
\begin{align}\label{H}
 \mathcal{H}^{}_0=\sum_n\epsilon^{}_n|n\rangle\langle n|-\sum_{\langle nm\rangle}(J^{}_{nm}|n\rangle\langle m|+{\rm h.c.})~.
\end{align}
For the special case of one-dimensional rings, which contain $N$ single-level
quantum dots, this becomes
\begin{align}\label{H1}
 \mathcal{H}^{}_0=\sum_n\bigl [\epsilon^{}_n|n\rangle\langle n|-(J^{}_{n,n+1}|n\rangle\langle n+1|+{\rm h.c.})\bigr ]~,
\end{align}
where  $n=N+1$ is identical to $n=1$. In some cases we shall
assume that the ring is penetrated by a magnetic flux $\Phi$
(measured in units of $\phi^{}_0/(2\pi)=\hbar c/e$). The
Aharonov-Bohm effect adds phases to the hopping terms,
$J^{}_{n,n+1}=J^{(0)}_{n,n+1}e^{i\phi^{}_{n,n+1}}$, where
$J^{(0)}_{n,n+1}$ is real and $\Phi=\sum^N_{n=1}\phi^{}_{n,n+1}$.
This flux generates  a circulating current around the loop, which  is discussed in Sec. IV. Additional comments on the results are given in Sec. V.

\section{Partial decoherence}

\subsection{General}

Generally, the time evolution of the reduced density matrix is given by
\begin{align}\label{rho1}
\rho^{}_{\ell\ell'}(t)&={\rm Tr}^{}_{\rm env}\langle\ell|\Psi(t)\rangle\langle\Psi(t)|\ell'\rangle\nonumber\\
&={\rm Tr}^{}_{\rm env}\langle\ell|e^{-i{\cal H}t}\Psi(0)\rangle\langle\Psi(0)e^{i{\cal H}t}|\ell'\rangle\ .
\end{align}
This expression becomes simple if the two energy eigenstates $|\ell\rangle$ and $|\ell'\rangle$ are both also eigenstates of ${\cal V}$,
 \begin{align}
 {\cal  V}|\ell\rangle=v^{}_{\ell}|\ell\rangle\ .\label{eigV}
 \end{align}
In this special case, Eq. (\ref{rho1}) can be written as\cite{16}
\begin{align}
\rho^{}_{\ell\ell'}(t)&={\rm Tr}^{}_{\rm env}\bigl [e^{-i{\cal H}^{}_{\ell}t}\langle\ell|\Psi(0)\rangle\langle\Psi(0)|\ell'\rangle e^{i{\cal H}^{}_{\ell'}t}\bigr ]\nonumber\\
&={\rm Tr}^{}_{\rm env}\bigl [e^{i{\cal H}^{}_{\ell'}t}e^{-i{\cal H}^{}_{\ell}t}\langle\ell|\Psi(0)\rangle\langle\Psi(0)|\ell'\rangle
\bigr ]\ ,\label{rho3}
\end{align}
where ${\cal H}^{}_\ell=\tilde{\epsilon}^{}_\ell{\bf 1}+{\cal H}^{}_{\rm env}+v^{}_\ell{\cal U}^{}_{\rm env}$ acts only on the environment states, and ${\bf 1}$ is the unit operator there.
For $\ell=\ell'$, one has $e^{i{\cal H}^{}_{\ell'}t}e^{-i{\cal H}^{}_{\ell}t}={\bf 1}$, and thus
\begin{align}
\rho^{}_{\ell\ell}(t)={\rm Tr}^{}_{\rm env}\langle\ell|\Psi(0)\rangle\langle\Psi(0)|\ell\rangle
\equiv\rho^{}_{\ell\ell}(0)\ .\label{rho2}
 \end{align}
 Since
 \begin{align}
 \rho^{}_{\ell\ell}=\sum_{nm}\langle\ell|n\rangle\rho^{}_{nm}\langle m|\ell\rangle\ ,
 \end{align}
 Eq. (\ref{rho2}) represents a combination of the reduced density matrix elements (in the site representation) which remains constant at all times, preserving some of the initial information.
 Furthermore, if $v^{}_\ell=v^{}_{\ell'}$ then Eq. (\ref{rho3}) reproduces Eq. (\ref{rabi}), so that the whole Hilbert sub-space spanned by the states $|\ell\rangle$ and $|\ell'\rangle$ remains entangled\cite{16}  and decoherence free.

 The number of time-independent invariants thus depends on how many basis states $|\ell\rangle$ are also eigenstates of ${\cal V}$. The maximal possible number is $N$. In that case, Eq. (\ref{eigV}) holds for {\it all} $\ell$, which implies the commutation $[{\cal H}^{}_0,{\cal V}]=0$ and thus adiabatic decoherence. It follows that Eq. (\ref{rho2}) also holds for all $\ell$. Since normalization requires the relation $\sum_\ell\rho^{}_{\ell\ell}=1$ (we allow only one electron in the system, and this electron stays there forever), we end up with $N-1$ independent relations among the density matrix elements $\rho^{}_{nm}(t)$, which remain invariant at all times.
 However, the number of time-invariants can have any value between $0$ and $N-1$, as we demonstrate in the following sections.

 As explained above, adiabatic decoherence implies that
 all the diagonal reduced density matrix elements remain time-invariant, while all the off-diagonal elements usually decay to zero. Therefore, this situation has been called "{\it pure dephasing}". Even in this  simpler case of decoherence,
a decay of off-diagonal matrix elements has been proven  only for
specific models with specific approximations.\cite{16,privman} It has also been found in our calculations for non-adiabatic decoherence, where we replaced the environment by a fluctuator which is described by telegraph noise.\cite{condmat}
It would be useful to have a general analysis of this decay, including criteria for when it occurs.
 In this paper we proceed with the assumption that the off-diagonal elements (in the energy basis) indeed decay to zero.

For non-adiabatic decoherence, the number of time-independent diagonal reduced density matrix elements may be smaller than $N$. When $\rho^{}_{\ell\ell}$ is not conserved, the system and the environment states remain entangled. This probably yields relaxation of these matrix elements into themalization. However, we are not aware of a general analysis of this process. Below we concentrate on the conserved diagonal elements.

\subsection{An alternative approach}

As we have shown, $\rho^{}_{\ell\ell}$ is time-invariant if the state $|\ell\rangle$ is an eigenstate of ${\cal V}$,
Eq. (\ref{eigV}). One way to proceed is thus to find the states $|\ell\rangle$, and then apply Eqs. (\ref{eigV})  to find conditions for partial decoherence. We employ this approach in Sec. IIIA below. We now present an alternatice approach, which avoids the explicit identification of the states $|\ell\rangle$. This approach is shown to be more useful in the example presented in Sec. IIIB.

Consider an hermitian operator ${\cal O}$,
which acts only on the system, and assume that this operator
commutes separately with {\it both} ${\cal H}^{}_0$ and ${\cal
V}$:
\begin{align}
[{\cal H}^{}_0,{\cal O}]=0\ ,\ \ \ [{\cal V},{\cal O}]=0\ .\label{commm}
\end{align}
Since the Heisenberg equation of motion of this operator has the
form
\begin{align}
\dot{\cal O}=i[{\cal H}^{}_0+{\cal V}{\cal U}^{}_{\rm env},{\cal
O}]=0\ ,\label{eom}
\end{align}
 it follows that ${\cal O}$ is independent of
time, and therefore also in the Schr\"{o}dinger representation one has
\begin{align}\label{Vav}
\langle\Psi(t)|{\cal O}|\Psi(t)\rangle=\langle\Psi(0)|{\cal
O}|\Psi(0)\rangle={\rm const.}\ ,
\end{align}
independent of time. Taking the trace over the states of the environment then yields
\begin{align}
&\langle\Psi(t)|{\cal O}|\Psi(t)\rangle={\rm Tr}[|\Psi(t)\rangle\langle\Psi(t)|{\cal O}]\nonumber\\
&={\rm tr}\{{\rm Tr}^{}_{\rm env}[|\Psi(t)\rangle\langle\Psi(t)|{\cal O}]\}
\equiv {\rm tr}\bigl [\rho(t){\cal O}\bigr ]\nonumber\\
&\equiv \sum_{mn}\rho^{}_{mn}(t){\cal O}^{}_{nm}=\sum_{mn}\rho^{}_{mn}(0){\cal O}^{}_{nm}\ ,\label{trace}
\end{align}
 where ${\rm Tr}$ and ${\rm tr}$ denote traces over {\it all} the states of the system+environment and over the states of the system only. The last equality  gives a relation between the elements of the reduced density matrix, $\rho^{}_{mn}(t)$,
 which must hold at all times, implying partial decoherence.
 As we show below, the existence of an operator ${\cal O}$ which
 satisfies the above conditions depends on the relations between
 the operators ${\cal H}^{}_0$ and ${\cal V}$.

 We start with the condition $[{\cal H}^{}_0,{\cal O}]=0$. This implies that we can diagonalize ${\cal H}^{}_0$ and ${\cal O}$
 simultaneously,  i.e. \begin{align}
 {\cal O}| \ell\rangle=o^{}_\ell |\ell\rangle\ ,
 \end{align}
with $N$ arbitrary real eigenvalues $\{o^{}_\ell\}$.  The second
requirement in Eq. (\ref{commm}), $[{\cal V},{\cal O}]=0$, then
implies
\begin{align}
\langle\ell'|{\cal V}|\ell\rangle(o^{}_{\ell'}-o^{}_\ell)=0\
.\label{oo}
\end{align}
If all the off-diagonal elements of ${\cal V}$ are not zero, then
all the $o^{}_\ell$'s are equal to each other, and the only
solution for ${\cal O}$ is proportional to the $N\times N$ unit
matrix. Equation (\ref{trace}) then implies that
$\sum_\ell\rho^{}_{\ell\ell}(t)$ is a constant, but this sum is
anyway equal to 1 due to the normalization.

To have at least one diagonal element $o^{}_{\ell}$ which differs
from all the others, we need at least one column of off-diagonal
matrix elements of ${\cal V}$ to vanish,
\begin{align}
\langle\ell'|{\cal V}|\ell\rangle=0\ {\rm for\ a\ given}\ \ell\  {\rm and\ for}\ all\  \ell'\ne
\ell\ .\label{lVl}
 \end{align}
 This condition is equivalent to the equality (\ref{eigV}),
  so that
 this particular state $|\ell\rangle$ is also an eigenstate of ${\cal V}$.
 In this case, we can choose $o^{}_{\ell'}=\delta^{}_{\ell\ell'}$, i.e. ${\cal O}=|\ell\rangle\langle\ell|$, and  Eq. (\ref{trace}) implies that
$\rho^{}_{\ell\ell}(t)=\rho^{}_{\ell\ell}(0)$, as in Eq. (\ref{rho2}).
However, other eigenstates of ${\cal H}^{}_0$ need not obey Eq. (\ref{eigV}).

So far, we have reproduced here the same results found in Sec. IIA. However, below we show an example in which this approach is easier to use.

\subsection{Adiabatic decoherence}

For $N=2$, ${\cal V}$ has only one off-diagonal matrix element, $\langle 1|{\cal V}|2\rangle$ (in the basis of the system's energy eigenstates). Therefore, the states $|\ell\rangle$ become eigenstates of ${\cal V}$ only if $\langle 1|{\cal V}|2\rangle=0$. In this case, ${\cal V}$ becomes diagonal, and it commutes with ${\cal H}^{}_0$, which implies adiabatic decoherence.
For $N=2$, one can write ${\cal H}^{}_0=a{\bf I}+{\bf b}\cdot\sig$
and ${\cal V}=c{\bf I}+{\bf d}\cdot\sig$, where ${\bf I}$ is the
$2\times 2$ unit matrix, $\sig$ is the vector of the three Pauli
matrices and the coefficients are all real (for hermiticity). The
equation $[{\cal H}^{}_0,{\cal V}]=0$ then implies that $[{\bf
b}\times{\bf d}]=0$, and therefore - apart from a trivial constant
shift in $a$ (to the new value $c/K$), equivalent to a shift in the zero of the energy of
the system - one must have ${\cal V}=K{\cal H}^{}_0$, where $K$ is
a real number. For a given operator ${\cal V}$, one can use gate voltages to tune the
site and hopping energies of ${\cal H}^{}_0$ (i.e. the three components of ${\bf b}$, or the energy difference  $\epsilon^{}_2-\epsilon^{}_1$ and the complex coupling $J^{}_{12}$ in Eq. (\ref{H})) in order to obey this
condition. \cite{prb}

For $N=2$, adiabatic decoherence requires the relation ${\cal V}=K{\cal H}^{}_0$. Of course, a relation like ${\cal V}=K{\cal H}^{}_0$ implies adiabatic decoherence also for larger $N$. However, for $N>2$ one can have adiabatic decoherence while ${\cal V}$ is not proportional to ${\cal H}^{}_0$ (but still commutes with it). Again, we predict that all the diagonal reduced density matrix elements remain time-invariant.
In both cases one might encounter situations in which some eigenvalues of ${\cal V}$ are degenerate (see below).  In that case, the corresponding off-diagonal matrix elements of $\rho$ will exhibit Rabi oscillations, without any dephasing. The whole subspace of the degenerate states is then decoherence free.

\section{Examples of non-adiabatic partial decoherence}

The environment can couple to the system in many ways. Here we concentrate on the special cases in which the coupling of the system to the environment occurs only via a sub-system, of dimension $M$. For example, in the case $N=3$ we assume that only the `control register' $|3\rangle$ is coupled to the environment, hence $M=1$. When the environment couples only to the sites $n=N-M+1,~\ldots,N$, this requirement becomes
\begin{align}
\langle\ell'|{\cal V}|\ell\rangle=\sum_{n,m=N-M+1}^N\langle \ell'|n\rangle\langle n|{\cal V}|m\rangle\langle m|\ell\rangle=0\ ,\label{ss}
\end{align}
for a given $\ell$ and all $\ell\ne \ell'$.

\subsection{$M=1$}

For $M=1$, the environment couples only to the state $|N\rangle$,
and the right hand side of Eq. (\ref{ss}) contains only one term, with $m=n=N$. This describes the examples mentioned after Eq. (\ref{HHH}). In this case, ${\cal V}$ clearly does not commute with ${\cal H}^{}_0$.
Since $\langle N|{\cal V}|N\rangle\ne 0$, Eq. (\ref{ss}) holds only when we have $\langle N|
\ell\rangle=0$.  Indeed, if an eigenstate $|\ell\rangle$ of ${\cal
H}^{}_0$ is orthogonal to the state $|N\rangle$, then ${\cal
V}|\ell\rangle=|N\rangle\langle N|{\cal V}|N\rangle\langle N|\ell\rangle=0$ and the time evolution of this eigenstate is
not affected by the environment. Such an eigenstate then belongs
to the {\it decoherence-free subspace} of the Hilbert space of the
system.

Consider now the Hamiltonian (\ref{H}), and denote $|\ell\rangle=\sum^{}_n\psi^{}_n|n\rangle$, with $\psi^{}_n=\langle n|\ell\rangle$.
The Schr\"{o}dinger equation for $|\ell\rangle$ becomes
\begin{align}
(\tilde{\epsilon}^{}_{\ell}-\epsilon^{}_n)\psi^{}_n=-\sum_m J^{}_{nm}\psi^{}_{m}\ ,
\end{align}
for $n=1,\ \ldots,\ N$. To achieve partial decoherence, we need states with $\psi^{}_N=0$. Therefore, the equation for $n=N$ becomes
\begin{align}
\sum_m J^{}_{Nm}\psi^{}_{m}=0\ .\label{eqN}
\end{align}
The other $N-1$ equations involve only the subsystem which contains the remaining $N-1$ sites, without the bonds connecting them to the site $N$. Solving those equations yields the $\psi^{}_m$'s, and substituting them into Eq. (\ref{eqN}) then gives a relation among the $J^{}_{Nm}$'s which must be obeyed in order to achieve partial decoherence.

Consider now the ring with $N$ sites, in which only the state $|N\rangle$ couples to the environment. Using gauge invariance, we place the whole flux on the bond $(N1)$, i.e.
$\phi^{}_{n,n+1}=\Phi \delta^{}_{n,N}$. Equation (\ref{eqN}) becomes
\begin{align}
J^{(0)}_{N,1}e^{i\Phi}\psi^{}_1+J^{(0)}_{N,N-1}\psi^{}_{N-1}=0\ .\label{psi}
\end{align}
The remaining $N-1$ equations, which  now  represent an open chain with $N-1$ sites and with vanishing boundary conditions, have only real coefficients $J_{n,n+1}$, and therefore they always have  solutions with {\it real} $\psi^{}_n$'s. Substituting back into Eq. (\ref{psi}), this implies that we can have partial decoherence only if $z=e^{i\Phi}=\pm 1$, i.e. if $\Phi$ is equal to zero or to $\pi$. Furthermore, given the $\psi^{}_n$'s which result from these $N-1$ equations, Eq. (\ref{psi}) also imposes a particular ratio between $J^{(0)}_{N,1}$ and $J^{(0)}_{N,N-1}$,
\begin{align}
J^{(0)}_{N,1}/J^{(0)}_{N,N-1}=-z\psi^{}_{N-1}/\psi^{}_1\ .
\label{JJ}
\end{align}

For the specially symmetric case, with $\epsilon^{}_n\equiv 0$ and $J^{(0)}_{n,n+1}\equiv J$, the Hamiltonian of the open chain with $N-1$ sites is symmetric under $n \leftrightarrow N-n$. Therefore, its eigenstates should also obey $\psi^{}_n=\pm\psi^{}_{N-n}$ with the two signs representing symmetric and anti-symmetric states. In particular, one has $\psi^{}_1=\pm\psi^{}_{N-1}$. Having assumed that all the $J$'s are equal, Eq. (\ref{JJ}) implies that one has partial decoherence for the antisymmetric states if $z=1$ and for the symmetric states if $z=-1$.

Since adiabatic decoherence requires that $[{\cal H}^{}_0,{\cal V}]=0$, and since now $\langle 1|[{\cal H}^{}_0,{\cal V}]|N\rangle=J^{}_{1,N}\ne 0$, it is clear that we cannot conserve {\it all} the diagonal $\rho^{}_{\ell\ell}$'s.
Therefore, there is no partial decoherence for this case when $N=2$, where the only possible partial decoherence is adiabatic.
For $N=3$, the symmetric and anti-symmetric cases relate to the `bonding' and `anti-bonding' states of the qubit, $(|1\rangle\pm|2\rangle)/\sqrt{2}$. Partial decoherence is obtained for the state $|\ell\rangle=(|1\rangle-z|2\rangle)/\sqrt{2}$, and therefore
\begin{align}
\rho^{}_{\ell\ell}(t)=\rho_{11}(t)+\rho^{}_{22}(t)-2z{\rm Re}\rho^{}_{12}(t)={\rm const.}
\end{align}\label{N33}
or, equivalently,
\begin{align}
\rho_{33}(t)+2z{\rm Re}\rho^{}_{12}(t)={\rm const.}\label{N3inv}
\end{align}
Indeed, these relations were found to hold in an explicit calculation of the time evolution of the $N=3$ ring, using the telegraph noise model for the fluctuations in ${\cal U}^{}_{\rm env}$.\cite{condmat}

The same result (\ref{N3inv}) would follow from the method presented in Sec. IID. It is easy to check that for the symmetric ring with $N=3$, the operator
\begin{align}
{\cal O}=|3\rangle\langle 3|+|1\rangle\langle 2|+|2\rangle\langle 1|
\end{align}
commutes with both ${\cal H}^{}_0$ and ${\cal V}$, and therefore Eq. (\ref{trace}) yields Eq. (\ref{N3inv}).

For symmetric rings with larger $N$, the solutions within the open chain are $\psi_n=\sin(\pi\ell n/N)\sqrt{2/N}$, $\ell=1,~2,~\ldots,~N-1$, with $\psi^{}_n=(-1)^{\ell-1}\psi^{}_{N-n}$.   Therefore, there are $(N-1)/2$ (or $N/2$) symmetric and $(N-1)/2$ [or $(N-2)/2$] antisymmetric states when $N$ is odd (or even). The diagonal elements of $\rho$ corresponding to each symmetric (or antisymmetric) state will remain constant in time for $z=-1$ (or $z=1$). These correspond to
 \begin{align}
 \rho^{}_{\ell\ell}=\frac{2}{N}\sum_{n,m}\sin\Bigl (\frac{\pi\ell n}{N}\Bigr )\sin\Bigl (\frac{\pi\ell m}{N}\Bigr )\rho^{}_{nm}(t)={\rm const.}
 \end{align}
 For example, for $N=4$ and $z=1$ we have
 \begin{align}
 \rho^{}_{11}+\rho^{}_{33}-2{\rm Re}\rho^{}_{13}={\rm const.}\ ,
 \end{align}
 while for $z=-1$ we have {\it two} invariants,
 \begin{align}
 \rho^{}_{11}+2\rho^{}_{22}+\rho^{}_{33}+2{\rm Re}\rho^{}_{13}\pm 2\sqrt{2}{\rm Re}(\rho^{}_{12}+\rho^{}_{23})={\rm const.}
 \end{align}

\subsection{$M>1$}

For $M>1$, it is easier to follow the method outlined in Sec. IIB.
 We  now split
 the Hilbert space of the system into two subspaces, $A$ and $B$, of dimensions $N-M$ and $M$. In the basis of the sites, the bare initial Hamiltonian is
\begin{align}
{\cal H}^{}_0=\left [\begin{array}{cc}h^{}_{AA} & h^{}_{AB} \\
 h^{}_{BA} & h^{}_{BB}\end{array}\right ]\ ,\label{H00}
 \end{align}
 and the interaction is
 \begin{align}
{\cal V}=\left [\begin{array}{cc}0 & 0 \\
 0& v^{}_{BB}\end{array}\right ]\ .\label{VV}
 \end{align}
 Writing ${\cal O}$ in the form
\begin{align}
 {\cal O}=\left [\begin{array}{cc}o^{}_{AA} & o^{}_{AB} \\
 o^{}_{BA} & o^{}_{BB}\end{array}\right ]\ ,\label{O}
 \end{align}
 the commutator with ${\cal V}$ is
 \begin{align}[{\cal V},{\cal O}]=\left [\begin{array}{cc}0 & -o^{}_{AB}v^{}_{BB} \\
 v^{}_{BB}o^{}_{BA} & [v^{}_{BB},o^{}_{BB}]\end{array}\right ]=0\ .\label{VO}
 \end{align}
 The solution of the equation $v^{}_{BB}o^{}_{BA}=0$ depends on the determinant of $v^{}_{BB}$. In our case we assume $\det[v^{}_{BB}]\ne 0$, and then $o^{}_{AB}=0$, for any $M$.

  We next consider the commutator
 \begin{align}[{\cal H},{\cal O}]=\left [\begin{array}{cc}[h^{}_{AA},o^{}_{AA}] & h^{}_{AB}o^{}_{BB}-o^{}_{AA}h^{}_{AB} \\
 h^{}_{BA}o^{}_{AA}-o^{}_{BB}h^{}_{BA} & [h^{}_{BB},o^{}_{BB}]\end{array}\right ]\ ,\label{HO}
 \end{align}
where we set $o^{}_{AB}=0$.
It is now convenient to diagonalize separately the Hamiltonians of the two decoupled sub-spaces,
\begin{align}
h^{}_{AA}|\alpha\rangle=\epsilon^{}_\alpha|\alpha\rangle\ ,\ \ \ h^{}_{BB}|\beta\rangle=\epsilon^{}_\beta|\beta\rangle\ .
\end{align}
The vanishing of the diagonal parts in Eq. (\ref{HO}) implies that these eigenstates can also be chosen to be eigenstates of ${\cal O}^{}_{AA}$ and ${\cal O}^{}_{BB}$,
\begin{align}
o^{}_{AA}=\sum_\alpha o^{}_\alpha|\alpha\rangle\langle\alpha|\ ,\ \ \ o^{}_{BB}=\sum^{}_\beta o^{}_\beta|\beta\rangle\langle\beta|\ .
\end{align}
Equation (\ref{VO}) now requires the relation
\begin{align}
\langle\beta|v^{}_{BB}|\beta'\rangle(o^{}_\beta-o^{}_{\beta'})=0\ ,\label{vbb}
\end{align}
and the vanishing of the off-diagonal part of Eq. (\ref{HO}) requires the relation
\begin{align}
\langle\beta|h^{}_{BA}|\alpha\rangle(o^{}_\beta-o^{}_{\alpha})=0\ .\label{hab}
\end{align}
The first of these relations is similar to Eq. (\ref{oo}). Therefore, it will be satisfied if there is at least one column of zero off-diagonal matrix elements in the matrix of $v^{}_{BB}$, or if all the $o^{}_\beta$'s are equal to each other.

The second of these equations involves $h^{}_{BA}$. For the ring Hamiltonian (\ref{H1}), this matrix contains only two elements, namely $J^{}_{N,1}$ and $J^{}_{N-M+1,N-M}$. Therefore, to have non-trivial values of $o^{}_\alpha$ or of $o^{}_\beta$ we need to satisfy
\begin{align}
&\langle\beta|h^{}_{BA}|\alpha\rangle=\langle\beta|N\rangle J^{}_{N,1}\langle 1|\alpha\rangle \nonumber\\
&+\langle\beta|N-M+1\rangle J^{}_{N-M+1,N-M}\langle N-M|\alpha\rangle=0\ .\label{cucu}
\end{align}
Placing again the whole flux on the bond $(N1)$, the separate wave functions for the open chains in the $A$ and $B$ subspaces are real, and therefore Eq. (\ref{cucu}) can be satisfied only if $z=e^{i\Phi}=\pm 1$, together with a particular ratio between $J^{(0)}_{N,1}$ and $J^{(0)}_{N-M+1,N-M}$.

For the specially symmetric case $\epsilon^{}_n\equiv 0$ and $J^{(0)}_{n,n+1}\equiv J$, one has $\langle \beta|N\rangle=\pm\langle\beta|N-M+1\rangle$ and $\langle\alpha|1\rangle=\pm\langle\alpha|N-M\rangle$, and therefore the final invariants will again involve symmetric and antisymmetric eigenstates within each subspace. Specifically, for $N=3$ and $M=2$ we find
\begin{align}
\rho^{}_{11}+2z{\rm Re}\rho^{}_{23}={\rm const.}
\end{align}
For $N=4$ and $M=2$ one has two invariants,
 \begin{align}
 {\rm Re}[\rho^{}_{12}\pm \rho^{}_{34}]={\rm const.}
 \end{align}

 \section{circulating currents}

As mentioned, in some cases the asymptotic state of the system maintains a circulating current, which can be used for retrieving information on its initial state.
 The site state $|n\rangle$ is  an eigenstate of the position coordinate  of the electron.
 Therefore, the position operator can be written as
 \begin{align}
 \hat{n}=\sum_n n|n\rangle\langle n|\ .
 \end{align}
 The current (or the velocity) operator is thus
 \begin{align}\label{cur123}
 \hat{I}=d\hat{n}/dt=i[{\cal H},\hat{n}]\equiv \hat{I}^{}_s+\hat{I}^{}_e\ ,
 \end{align}
 where
 \begin{align}
 \hat{I}^{}_s=i[{\cal H}^{}_0,\hat{n}]\ ,\ \ \hat{I}^{}_e=i[{\cal V},\hat{n}]{\cal U}^{}_{\rm env}\ .
 \end{align}
 For the ring Hamiltonian (\ref{H1}), the first term becomes
 \begin{align}
 \hat{I}^{}_s=-i\sum_n\bigl (J^{}_{n,n+1}|n\rangle\langle n+1|-{\rm h.c.}\bigr )\ .\label{Is}
 \end{align}
 The quantum average of this current is
 \begin{align}\label{avI}
 \langle I^{}_s \rangle={\rm Tr}\bigl (\rho\hat{I}^{}_s\bigr )=2\sum_n{\rm Im}\bigl (\rho^{}_{n,n+1}J^{}_{n+1,n}\bigr )\ .
\end{align}

For simplicity, consider Eq. (\ref{H1}) in the fully symmetric case, $\epsilon^{}_n=0$
and $J^{(0)}_{n,n+1}=J$, and choose the gauge
$\phi^{}_{n,n+1}=\phi\equiv\Phi/N$. The eigenstates of ${\cal
H}^{}_0$ are \begin{align}
|\ell\rangle=\sum^{}_n e^{i2\pi \ell
n/N}|n\rangle/\sqrt{N}\ ,
\end{align}
  and $\tilde{\epsilon}^{}_\ell=-2J\cos(\phi+2\pi\ell/N)$, for $\ell=0,~1,\cdots,~N-1$. Therefore, the
conserved quantities are \begin{align}
\rho^{}_{\ell\ell}(t)=\rho^{}_{\ell\ell}(0)=\sum^{}_{mn}
e^{i(2\pi\ell/N) (m-n)}\rho^{}_{mn}(0)/N\ .\label{rll}
 \end{align}
 Assuming that the off-diagonal matrix elements $\rho^{}_{\ell\ell'}$ decay to zero at long times, and the asymptotic reduced density matrix in
the site representation becomes
\begin{align}
\rho^{}_{mn}(t\rightarrow\infty)=\sum^{}_\ell e^{i(2\pi
\ell/N)(n-m)}\rho^{}_{\ell\ell}(0)/N\ . \label{rnm}
\end{align}

For $N>2$, one may have degenerate states: Since $\tilde{\epsilon}^{}_\ell-\tilde{\epsilon}^{}_{\ell'}=4J\sin[\phi+\pi(\ell+\ell')/N]\sin[\pi(\ell-\ell')/N]$, these two states are degenerate if $\phi$ equals an integer multiple of $\pi$ and if $\ell+\ell'=N$. In these cases, $\rho^{}_{\ell\ell'}(t)$ is also time-invariant. Specifically, for the adiabatic decoherence of the $N=3$ symmetric ring with ${\cal V}=K{\cal H}^{}_0$, the whole subspace formed by the states $|\ell=1\rangle$ and $|\ell=2\rangle$ (i.e. the states of the qubit) is decoherence free.
However, if $v^{}_1\ne v^{}_2$ then only the diagonal matrix elements $\rho^{}_{11}$ and $\rho^{}_{22}$ are conserved.

For this special symmetric case, Eq. (\ref{rnm}) yields
\begin{align}
\langle I^{}_s\rangle=-\frac{2J}{N}\sum_\ell\rho^{}_{\ell\ell}\sin(\phi+2\pi\ell/N)=-\frac{1}{N}\frac{\partial}{\partial\phi}\langle{\cal H}^{}_0\rangle\ ,
 \end{align}
 which is equivalent to  the magnetic moment per site associated with the flux through the ring.

 The second term in Eq. (\ref{cur123}), $\hat{I}^{}_e$, represents current fluctuations due to the coupling to the environment. This term requires the average over the environment states of ${\cal U}^{}_{\rm env}$, and the commutator  $[{\cal V},\hat{n}]$. In the simplest adiabatic case, when ${\cal V}=K{\cal H}^{}_0$, this will result in a simple shift of the prefactor $J$ to $J^{}_0=J(1+K\langle{\cal U}^{}_{\rm env}\rangle)$. This was the expression used in Ref.$\ $\onlinecite{prb}. In the case of coupling at a single site, Sec. IIIA, ${\cal V}$ is diagonal in the site basis, hence $[{\cal V},\hat{n}]=0$. In  other cases, this term requires a separate analysis.

 For the special case $N=2$,  if we start with
the initial state (\ref{psi0}), we have
$\rho^{}_{\ell\ell}=[1+\sin(2\alpha)\cos(\pi\ell+\gamma)]/2$,  and thus $\rho^{}_{11}=\rho^{}_{22}=1/2,~~\rho^{}_{12}=\rho^{}_{21}=\sin(2\alpha)\cos\gamma/2$.  Therefore, $\langle I\rangle=J^{}_0\sin(2\alpha)\cos\gamma\sin\phi$.\cite{com}

For all the non-adiabatic cases discussed above, we found partial decoherence only for real $z=e^{i\Phi}$. Since the asymptotic reduced density matrix is also real, such values of $\Phi$ yield zero circulating current. However, this result is specific for {\it single rings}. As soon as the system contains more than one ring, the proof that $e^{i\Phi}$ must be real breaks down, and one can still have some circulating currents. We expect to address this more general case in another publication.

\section{conclusions}

We have shown that if a state $|\ell\rangle$ is simultaneously an eigenstate of both ${\cal H}^{}_0$ and ${\cal V}$ then $\rho^{}_{\ell\ell}(t)$ is time-invariant, allowing the retrieval of some initial quantum information at any time, despite the decoherence.
The number of such states $|\ell\rangle$ depends on the symmetry of the system, which can be modified by tuning the parameters in the system's Hamiltonian, and we have shown examples in which one finds either adiabatic decoherence (all diagonal reduced matrix elements are time-invariant) or non-adiabatic decoherence. In the latter case,
rings have an asymptotically vanishing circulating current.

So far, we have concentrated on the ideal case, when the Hamiltonian can be tuned exactly so that one achieves partial decoherence at all times. In Ref.$\ $\onlinecite{prb} we also discuss small deviations from such exact tuning. For the special cases treated there we found that such deviations cause a very slow decay of  the diagonal reduced density matrix elements, which would remain time-invariant for the exact tuning. The rate of the decay depended on these deviations, and the amplitudes were equal to the initial values of these matrix elements. Thus, one could still extract information on the initial state from these amplitudes. We expect similar phenomena also for the more general case discussed here. This would result from an appropriate expansion of Eq. (\ref{rho1}) in these deviations.
Similar slow decays are expected to arise due to additional weak coupling terms, involving other environments. We leave the analysis of such situations for future work.

We hope that this paper will stimulate some experimental tests of our results. We also hope that it will generate some general discussion on what happens to the reduced matrix elements which are not time-invariant.

\acknowledgements

SD and YT are grateful to BGU for the warm hospitality they received during
several visits, when this project was carried out. AA, SG and OEW
acknowledge  support from the ISF. AA and OEW also acknowledge support from the Albert Einstein Minerva Center for Theoretical
Physics at the Weizmann Institute of Science. YT acknowledges the support of the Funding Program for World-Leading Innovative R$\&$D on Science and Technology (FIRST).


\begin{thebibliography}{21}
\bibitem{com} In Ref.$\ $\onlinecite{prb}, the two sites were connected with 2 bonds, with the hopping terms $(J^{}_u+J^{}_d e^{i\Phi})|1\rangle\langle 2|+{\rm h.c.}$. In the present paper, we use a different gauge,
    $(J^{}_ue^{-i\phi}+J^{}_d e^{i\phi})|1\rangle\langle 2|+{\rm h.c.}$, with $J^{}_u=J^{}_d=J$ and $\phi=\Phi/2$. The transition between these two gauges requires the mapping of $|2\rangle$ to $e^{i\phi}|2\rangle$. This mapping requires also $\gamma\rightarrow\gamma+\phi$. For $J^{}_u=J^{}_d$, the final circulating currents found in the two papers agree with each other.
\end{thebibliography}
\end{document}